\documentstyle[amstex,12pt]{article}

\newtheorem{pro}{\large\sc Proposition}

\newtheorem{lem}{\large\sc Lemma}
\begin{document}
\vspace{1cm}
\title{Scaling Self--Similar Formulation of the   String Equations of the
Hermitian One--Matrix Model}
\author{
Manuel Ma\~nas\thanks{Research supported by British
Council's Fleming award ---postdoctoral MEC   fellowship GB92 00411668, and
postdoctoral
EC Human Capital and Mobility individual fellowship ERB40001GT922134}\\  The
Mathematical Institute\\
Oxford University\\ 24-29 St.Giles', Oxford OX1 3LB \\United Kingdom}
 \maketitle
\begin{abstract}
The string equation appearing in the double
scaling limit of the Hermitian one--matrix model, which corresponds to
a Galilean self--similar condition for the KdV hierarchy, is reformulated as a
scaling self--similar condition
for the  Ur--KdV hierarchy. A non--scaling limit analysis of the one--matrix
model has  led to the
  complexified NLS hierarchy and a string equation. We show that this
corresponds to the
Galilean self--similarity condition for the AKNS hierarchy and also  its
equivalence
to a scaling self--similar condition for the  Heisenberg ferromagnet hierarchy.
\end{abstract}

{\bf 0.} The Hermitian  one--matrix model has received much attention in recent
years as a non--perturbative formulation of string theory. In   \cite{bk} the
double scaling limit for the even potential case was used to show that the
specific heat   is a solution of the Korteweg--de Vries (KdV)  hierarchy that
satisfies an additional constraint, the so called string equation. In \cite{l}
it was prove
that this corresponds to invariance under Galilean transformations, see also
\cite{gm2}. The model is also relevant for topological gravity and for the
Witten--Kontsevich intersection theory of the moduli space of complex curves
\cite{wk}.

In \cite{bx} it was performed a non--scaling limit analysis of the the
Hermitian one--matrix model with general potential. Now, the specific
heat is the second conserved density of the
Ablowitz--Kaup--Newell--Segur (AKNS) hierarchy and the string
equation, as we shall show, corresponds to invariance under the
Galilean transformations of the hierarchy. The associated topological
field theory is close to the Witten's ${\Bbb C}P^1$ $\sigma$--model
coupled to topological gravity. Observe that the AKNS hierarchy is a
complexfied version of the Non--Linear Schr\"odinger (NLS) hierarchy.

The aim of this letter is to show that the Hermitian one--matrix model, which
  corresponds to Galilean self--similarity ---$L_{-1}$--Virasoro constraint---
with or without double scaling limit, can be formulated
as a $L_0$--Virasoro constraint, that is, as a scaling invariance condition. In
order to do this we need to introduce
new integrable hierarchies connected with the previous ones by  Miura type
transformations.

For the string equation of the double scaling limit of the Hermitian
one--matrix
model we introduce the fundamental Ur--KdV hierarchy, see \cite{w,mc} and
references therein.    And, as we shall see,
the Galilean self--similarity condition, i.e. the string equation, corresponds
in the Ur--KdV hierarchy to scaling self--similarity. For example, the solution
corresponding to the Witten--Kontsevich model when only the KdV equation is
taken into account, corresponds in the Ur--KdV equation to a linear fractional
transformation of a quotient of Airy functions
depending on a scaling invariant. Recall that the Airy equation is important in
Kontsevich's approach.

We consider as well the string equations of the non-scaling limit  of the
Hermitian one--matrix model \cite{bx}. We not only prove that these are
equivalent to the
Galilean self--similarity condition for the AKNS hierarchy but also that for
the associated Heisenberg ferromagnet hierarchy \cite{ft} this corresponds to
scaling self--similarity.

In the first section we introduce the Ur--KdV hierarchy. There one can find the
proof of the equivalence of the string equation of the double scaling limit of
the Hermitian one--matrix model with the scaling self--similar condition for
solutions of the Ur--KdV hierarchy. We also illustrate the correspondence with
the Galilean self--similar solution of the KdV equation and the appearence of
the Airy functions. For the scaling self--similar condition of the KdV
hierarchy we found its equivalent in the Ur--KdV hierarchy, giving as example
the Adler--Moser rational solutions.
We end the section presenting a general formula expressing the solution to the
Ur--KdV hierarchy
as a quotient of two $\tau$--functions for the KdV hierarchy.

In the next section we analyse the non--scaling limit case. First we prove that
the string equation of \cite{bx} corresponds to the Galilean self--similarity
condition for the AKNS hierarchy. Secondly, we introduce the complexified
version of the continuous spin chain with Heisenberg interaction, that is, the
Heisenberg
ferromagnet  hierarchy, showing that the string equation can be recasted as a
scaling self--similar condition for this hierarchy. To end   we look
to the scaling self--similar condition of the AKNS hierarchy and to the
corresponding condition in the Heisenberg ferromagnet hierarchy.

{\bf 1.} The Ur--KdV equation,  as named by Wilson \cite{w}, is the following
non--linear partial differential equation for a complex scalar field
$z$ depending on the complex variables $t_1,t_3$
\[
4 \partial_3z=\{ z,t_1\}\partial_1z
\]
where
\[
\{ z,t_1\}:=\frac{\partial_1^3z}{\partial_1 z}-\frac{3}{2}\left(
\frac{\partial_1^2z}{\partial_1z}\right)^2
\]
is the Schwartzian derivative \cite{f}. Here we have use the notation
$\partial/\partial t_1=:\partial_1$ and so on.

This equation is connected to the KdV equation. Given a solution $z$ to the
Ur--KdV equation then
\begin{equation}
u=\frac{1}{2}\{z,t_1\} \label{trans}
\end{equation}
satisfies the KdV equation
\[
4\partial_3u=\partial_3u+6u\partial_1u.
\]

The Ur--KdV equation is associated with the Krichever--Novikov equation
\[
4 \partial_3z=\{ z,t_1\}\partial_1z+\frac{4z^3-g_2z-g_3}{\partial_1z}
\]
which appears in the study of rank 2 and genus 1 solutions to the
KP equation,   \cite{kn}. It is also important in the
classification of scalar integrable equations of order 3, \cite{ss}. And can be
described in terms of certain elliptic
homogeneous spaces in analogy to the Landau--Lishfitz equation,
\cite{gm1}.

As is well known, the KdV equation has an infinite number of
symmetries that preserves the spectral properties of the
associated Schr\"odinger operator
\[
{\cal L}:=\partial_1^2 +u .
\]
Therefore, one comes to consider the KdV hierarchy, that can be
expressed in terms of the Gel'fand--Dickii potentials $R_n[u]$ (polynomials in
$u,\partial_1u,\partial_1^2u,\dots$)
which are the coefficients of an asymptotic expansion of the
resolvent $({\cal L}-\lambda)^{-1}$, \cite{gd}. The KdV
hierarchy is the following infinite set of compatible equations in the
variables ${\bf t}:=\{t_{2n+1}\}_{n\geq 0}$
\[
\partial_{2n+1}u=4\partial_1R_{n+1}[u].
\]
There is a corresponding Ur--KdV hierarchy
\begin{equation}
 \partial_{2n+1}z=2R_n\left[\frac{1}{2}\{ z,t_1\}\right]\partial_1
z,\label{evo}
\end{equation}
and any of its solutions gives through (\ref{trans}) a solution to
the KdV hierarchy. The Ur--KdV hierarchy has a remarkable property,
given a solution $z$ any
\[
\tilde z=\frac{az+b}{cz+d}
\]
is also a solution as long as $ab-cd=1$. Thus, the projective group
$PSL(2,\Bbb C)$ acts on the space of solutions to the hierarchy.

 Consider a solution $u$ of the KdV hierarchy, choose two independent functions
$\psi_1,\psi_2$ in the kernel of the Schr\"odinger operator $\cal L$ with
Wronskian equal to the unity. Hence
\[
{\cal L}\psi_1={\cal L}\psi_2=0,\;
\text{W}(\psi_1,\psi_2):=\psi_1\partial_1\psi_2-\psi_2\partial_1\psi_1=1.
\]
In addition we require  both $\psi_1,\psi_2$ to be in the kernel of the
evolution
operators
\begin{equation}
{\cal A}_{2n+1}=\partial_{2n+1}-2R_n\circ\partial_1+
(\partial_1 R_n).\label{evo1}
\end{equation}
 Thus, as one can show
\[
z=\frac{\psi_1}{\psi_2}
\]
satisfies the Ur--KdV hierarchy and is connected to $u$ through
(\ref{trans}). This can be considered as an inversion of (\ref{trans}).

Let us now consider the symmetries defined by
translations, scaling and
Galilean
transformations of the KdV hierarchy.
For the infinite set of
translational symmetries  we define
\[
\vartheta({\bf t}):={\bf t}+\mbox{\boldmath$\theta$},
\]
where
\[
\mbox{\boldmath$\theta$}:=\{\theta_{2n+1}\}_{n\geq 0}
\in{\Bbb C}^{\infty}.
\]
 If $u$ is a solution
to the hierarchy then
$\vartheta^{\ast}u$ is also a solution. For the scaling symmetry  ${\bf
t}\mapsto\varsigma_{\sigma}({\bf t})$ we define
\[
\varsigma_{\sigma}({\bf t})_{2n+1}:=
e^{(n+\frac{1}{2})\sigma}t_{2n+1}
\]
where $\sigma\in{\Bbb C}$. If $u$ is
a solution of the   KdV hierarchy then $e^{\sigma}\varsigma_{\sigma}^{\ast}u$
is a solution as well.
 The Galilean transformation  ${\bf t}\mapsto\gamma_b({\bf t})$ is given by
 \[
\gamma_b({\bf t})_{2n+1}:=
\sum_{m=0}^{\infty}
\binom{n+m+\frac{1}{2}}{m+\frac{1}{2}} b^m t_{2(n+m)+1},
\]
where we have used the binomial function that can be expressed in terms of the
Euler $\Gamma$--function as
\[
\binom{a}b:=\frac{\Gamma(a+1)}{\Gamma(b+1)\Gamma(a-b+1)}.
\]
  If $u$ is solution of the
KdV hierarchy then so is $\gamma_b^{\ast}u+b$.

The vector fields generating these symmetries  are
\[
\partial_{2n+1},n\geq 0,
\ \mbox{\boldmath$\varsigma$}
=\sum_{n\geq 0}(n+\frac{1}{2})t_{2n+1}\partial_{2n+1}, \
\mbox{\boldmath$\gamma$}=\sum_{n\geq 1}
(n+\frac{1}{2}) t_{2n+1}\partial_{2n-1},
\]
for translations, scaling and Galilean transformations respectively.
  Consider the following vector field
\[
X:=
\mbox{\boldmath$\vartheta$}+ \mbox{\boldmath$\gamma$},
\]
with
\[
\mbox{\boldmath$\vartheta$}:=\sum_{n\geq 0} \theta_{2n+1}
\partial_{2n+1}.
\]

Let us denote
\[
{\cal R}:=\sum_{n\geq 0}(n+\frac{1}{2})t_{2n+1}R_n,
\]
then, as one can show,
a solution $u$ of the  KdV hierarchy
is self--similar under the vector field $X$, i.e. $u$ remains invariant under
the symmetry transformation generated by $X$,
if and only if
it satisfies the  string equation
\begin{equation}
\sum_{n\geq 0}\theta_{2n+1}R_{n+1}+{\cal R}=\frac{c}{4}\label{sssecu}
\end{equation}
for some $c\in\Bbb C$. Notice that the self--similar solutions
under the vector field $X$ can be understood as Galilean self--similar
solutions
once we shift the times by $t_{2n+1}\mapsto t_{2n+1}+(n+1/2)^{-1}\theta_{2n-1}$
where $n\geq 1$. A shift in
$t_1$ changes the constant $c$. Therefore, any Galilean self--similar solution
of the KdV hierarchy is of the form
$u(t_1-c,t_3,\dots)$ where $u$ is a solution of the string
equation of the double scaling limit of the Hermitian one--matrix model with
even potentials \cite{bk,l,gm2}
\[
{\cal R}=0.
\]

In the Ur--KdV hierarchy there is no Galilean local symmetry, only the scaling
transformation
and the translations are local symmetries.
If $z$ is
a solution of the   Ur--KdV hierarchy then  $\vartheta^{\ast}z$ and
$\varsigma_{\sigma}^{\ast}z$
are  solutions as well.
  Now, we consider the   vector field
\[
Y:=
\widetilde{\boldsymbol{\vartheta}}+  \boldsymbol{\varsigma},
\]
with
\[
\widetilde{\mbox{\boldmath$\vartheta$}}:=-\frac{c}{2}\partial_1+\sum_{n\geq 1}
\theta_{2n-1}
\partial_{2n+1}.
\]

Using Eqs.(\ref{evo},\ref{evo1}) one can show
\begin{lem}
The following relations holds
\begin{eqnarray}
Y\psi_i+\frac{\psi_i}{4}&=&2({\cal R}+\sum_{n\geq
0}\theta_{2n+1}R_{n+1}-\frac{c}{4})
\partial_1\psi_i-\psi_i\partial_1({\cal R}+\sum_{n\geq
0}\theta_{2n+1}R_{n+1}-\frac{c}{4}),\label{ecup}\\
Yz&=&2({\cal R}+\sum_{n\geq 0}\theta_{2n+1}R_{n+1}-\frac{c}{4})\partial_1z.
\label{ecuz}
\end{eqnarray}
\end{lem}
{}From where one deduces
\begin{pro}
A solution $u$ to the KdV hierarchy satisfies the string equation
\[
{\cal R}+\sum_{n\geq 0}\theta_{2n+1}R_{n+1}=\frac{c}{4}
\]
  only if the corresponding solution $z$ to the Ur--KdV hierarchy satisfies
\[
Yz=0.
\]
And if $Yz=0$ then the corresponding solution $u$ to the KdV hierarchy either
satisfies the
string equation or $u=0$.
\end{pro}
{\bf Proof:}
If the string equation is satisfied then Eq.(\ref{ecup}) imply that
\[
Y\psi_i+\frac{\psi_i}{4}=0,
\]
thus
\[
Yz=Y(\frac{\psi_1}{\psi_2})=0.
\]
This proves the only if part. Now, if $Yz=0$  then Eq.(\ref{ecuz}) gives that
either the string equation holds
or $\partial_1z=0$, so that $z=z_0\in\Bbb C$ and $u=0$.$\Box$

 We arrive to the conclusion that given a non--constant self--similar solution
$z$ of the Ur--KdV hierarchy under the action of the vector field $Y$
---generating scaling transformations in shifted coordinates---
then the associated solution of the KdV hierarchy by means of
(\ref{trans}) is self--similar under the action of the vector
field $X$ and viceversa.

As an example we consider the Galilean self--similar solution of the KdV
equation
\[
u=-\frac{2t_1}{3t_3}.
\]
 The corresponding scaling self--similar solution of the Ur--KdV equation is
\[
z(t_1,t_3)=\frac{a\text{Ai}(\zeta)+b\text{Bi}(\zeta)}{c\text{Ai}(\zeta)+d\text{Bi}(\zeta)},\; ad-bc=1
\]
being Ai and Bi the standard Airy functions \cite{sf} and
\[
\zeta^3:=\frac{2}{3}\frac{t_1^3}{ t_3}.
\]

The solution $u$ when extended to all the $t_{2n+1}$ and once
the shift $t_3\mapsto t_3+3/2$ is performed is the one considered by Kac and
Schwarz \cite{ks} and also the one associated to the Witten--Kontsevich
model for the intersection theory of the moduli space of complex curves
\cite{wk}. Observe the appearance of the Airy functions,
essential in the Kontsevich approach, in the Ur--KdV context.

To end this section we shall give the string equation
for the Ur--KdV hierarchy corresponding to the scaling self--similar condition
to the
KdV equation, i.e. to 2D--stable gravity \cite{d}. One can easily show
that the following relation holds
\[
e^{\sigma}\varsigma_{\sigma}^*u=\frac{1}{2}\{
\varsigma_{\sigma}^*z,t_1\}.
\]
Thus, if we want $u$ to be scaling self--similar, we arrive
to the condition
\[
\{\varsigma_{\sigma}^*z,t_1\}=\{z,t_1\}
\]
so that
\[
\varsigma_{\sigma}^*z=\frac{(\cosh(\sigma\rho)+A_0\frac{\sinh(\sigma\rho)}{\rho}) z+A_+\frac{\sinh(\sigma\rho)}{\rho}}
{(A_-\frac{\sinh(\sigma\rho)}{\rho})z+\cosh(\sigma\rho)-A_0\frac{\sinh(\sigma\rho)}{\rho}},
\]
 with $A=A_+E+A_0H+A_-F\in{\frak sl}(2,\Bbb C)$ and $\rho=-{\rm det}A$.
Hence, the scaling self--similarity condition for the KdV
hierarchy can be recasted as scaling self--similarity of the Ur--KdV hierarchy
modulo the global $PSL(2,\Bbb C)$--gauge invariance.

 \begin{pro}
A solution $u$ of the KdV hierarchy is scaling self--similar if and only if the
corresponding solution $z$ to the Ur--KdV hierarchy satisfies the string
equation
\[
\boldsymbol{\varsigma}z=-A_-z^2+2A_0z+A_+.
\]
for some $A_+,A_0,A_-\in\Bbb C$.
\end{pro}

When $A=0$ we recover the Galilean case already exposed. Another example is
$A=H/4$, then $\varsigma_{\sigma}^*\partial_1z=\partial_1z$. The function
\[
w=\frac{1}{2}\ln(\partial_1z),
\]
is a solution of the potential modified KdV hierarchy self--similar under
scaling transformations. As was shown in \cite{mg} this
corresponds to the double scaling limit of the symmetric unitary one--matrix
model with no boundary terms. We see that this
sector of the double scaling limit  of the one--matrix model  can be encoded
with the Hermitian one with the
aid of the Ur--KdV hierarchy.

As an illustration let us consider the rational solutions of the KdV
hierarchy that vanishes when $t_1\rightarrow\pm\infty$, \cite{am}. These are
self--similar solutions under scaling transformations. The rational solution
$u_n$ is characterized by $u_n(t_1,0,0,\cdots)=n(n+1)/t_1^2$ where $n\in{\Bbb
N}\cup\{0\}$. One has the expression $u_n=2\partial_1^2\ln\Theta_n$, where
$\Theta_n$ is a polynomial in ${\bf t}$ of degree $n(n+1)/2$
($\text{deg}\,t_{2n+1}=2n+1$) and can be
considered as a theta function for the rational curve $\mu^2=
\lambda^{2n+1}$, they are $\tau$--functions. For the corresponding
Schr\"odinger operator ${\cal L}_n=\partial_1^2+u_n$ one has the kernel
$\text{Ker}{\cal L}_n={\Bbb C}\{\Theta_{n+1}/\Theta_n,\Theta_{n-1}/\Theta_n\}$,
so that
\[
z=\frac{a\Theta_{n+1}+b\Theta_{n-1}}{c\Theta_{n+1}+d\Theta_{n-1}}
\]
is a solution of the Ur--KdV hierarchy. For example
$z_n:=\Theta_{n+1}/\Theta_{n-1}$ satisfies
$\varsigma_{\sigma}^*z_n=e^{(n+1/2)\sigma}z_n$.

The possibility of expressing $z$ as a quotient of $\tau$--functions for the
KdV hierarchy is true in general and not only for the rational case considered
above. Let $u_0=2\partial^2_1\ln\tau_0$ be a solution of the KdV hierarchy.
Consider the expressions $u_0=\partial_1v_+-v_+^2=-(\partial_1v_-+v_-^2)$ where
$v_+$ and
$v_-$ are solutions to the modified KdV hierarchy. Then
$u_0$ is a B\"acklund transformation of
$u_-=\partial_1v_--v_-^2=2\partial_1^2\ln\tau_-$, solution of the
KdV hierarchy, and generates the solution
$u_+=-(\partial_1v_++v_+^2)=2\partial_1^2\ln\tau_+$.
 Then, the corresponding solution of the Ur--KdV hierarchy  is of the form
$z=(a\tau_++b\tau_-)/(c\tau_++d\tau_-)$. Connected with this see \cite{mc}.

{\bf 2.}
 In this section we shall show that the string equation found for the Hermitian
one--matrix model in \cite{bx} is the Galilean self--similarity condition for
the AKNS hierarchy and then we shall prove that this is equivalent to the
scaling self--similarity condition for the associated Heisenberg ferromagnet
hierarchy.

The AKNS  hierarchy for $p,q$, functions depending on ${\bf t}=\{t_n\}_{n\geq
0}$ is the following collection of
compatible equations
\begin{equation}
\begin{cases}
\partial_np= 2p_{n+1},\\
\partial_nq=-2q_{n+1},
\end{cases}\label{akns}
\end{equation}
where $ n\geq 0$,
 $\partial_n:=\partial/\partial t_n$
and $p_n,q_n$ and $h_n$ are defined recursively by the relations
\begin{eqnarray*}
&&p_n=\frac{1}{2}\partial_1p_{n-1}+ph_{n-1},\\
&&q_n=-\frac{1}{2}\partial_1q_{n-1}+qh_{n-1},\\
&&\partial_1h_n=pq_n-qp_n,\;\; n\geq 1
\end{eqnarray*}
with the initial data
$
 p_0=q_0=0,\, h_0=1$.

The $n=0$ flow is   usually not considered in the standard AKNS hierarchy, but
its inclusion will prove  convenient.  The equations for that flow are
$
\partial_0p=2p,
\partial_0q=-2q,
$
which means that
$
p(t_0,t_1,\dots)=\exp(2t_0)\tilde p(t_1,\dots),
q(t_0,t_1,\dots)=\exp(-2t_0)\tilde q(t_1,\dots)$.
 The functions $(\tilde p,\tilde q)$ satisfy the standard AKNS hierarchy, and
this $t_0$--flow reflects the fact that given a solution  $(\tilde p,\tilde q)$
to the standard AKNS hierarchy ($n> 0$) then  $(e^c\tilde p,e^{-c}\tilde q)$ is
a solution as well for any $c\in{\Bbb C}$.
The $n=2$ flow is
\[
\begin{cases}
2\partial_2p=\partial_1^2p-2p^2q,\\
2\partial_2q=-\partial_1^2q+2pq^2.
\end{cases}
\]
Notice that the real reduction $q=\mp p^{\ast}$ and $t_n\mapsto it_n$ produces
the $\text{NLS}^{\pm}$ hierarchy for which the $t_2$--flow is
$2i\partial_2 p=-\partial_1^2p\pm 2|p|^2p$, the $\text{NLS}^{\pm}$ equation.

 Let us now describe the local symmetries of the AKNS hierarchy. First we have
the shifts in the time variables.
  Let $\vartheta$ be
\[
\vartheta({\bf t}):={\bf t}+\mbox{\boldmath$\theta$},
\]
the action of translations, where
\[
\mbox{\boldmath$\theta$}:=\{\theta_n\}_{n\geq 0}
\in{\Bbb C}^{\infty},
\]
are the shifts of the time variables.
 If $(p,q)$ is a solution
to the AKNS hierarchy  then so is $(\vartheta^{\ast}p,\vartheta^{\ast}q)$.

 The Galilean  transformation  ${\bf t}\mapsto\gamma_b({\bf t})$ is given by
\[
\gamma_b({\bf t})_n:=
\sum_{m\geq 0}
\binom{n+m}m
b^mt_{n+m}
\]
where $b\in{\Bbb C}$.
 The scaling transformation ${\bf t}\mapsto\varsigma_{\sigma}({\bf t})$ is
represented by the relations
\[
\varsigma_{\sigma}({\bf t})_n:=e^{n\sigma}t_n
\]
where $\sigma\in{\Bbb C}$.
 If $(p,q)$ is a
 solution of the AKNS hierarchy  then so are
$(\gamma_b^{\ast}p,\gamma_b^{\ast}q)$ and
$(e^{\sigma}\varsigma_{\sigma}^{\ast} p,e^{\sigma}\varsigma_{\sigma}^{\ast}q)$.

Notice that for the corresponding solutions $(\tilde p,\tilde q)$ of the
standard AKNS hierarchy  the Galilean action is
$(\exp(2t(a))\gamma_a^{\ast}\tilde p,(\exp(-2t(a))\gamma_a^{\ast}\tilde q)$,
the exponential factors are a result of the flow in $t_0$ induced by the
Galilean transformation.
The related fundamental vector fields, infinitesimal generators of the
action of translation, Galilean and scaling   transformations are
given by
\[
\partial_n,\;n\geq 0,
\,\;  \mbox{\boldmath$\gamma$}=\sum_{n\geq 0}(n+1)
t_{n+1}\partial_n,\,\;\mbox{\boldmath$\varsigma$}
=\sum_{n\geq 1}n t_n\partial_n,
\]
respectively.

Consider the vector field
\[
X:=\boldsymbol{\gamma}+\boldsymbol{\vartheta},
\]
with
\[
\boldsymbol{\vartheta}:=\sum_{n\geq 0}\theta_n\partial_n.
\]

Then, $(p,q)$ is a self--similar solution  under the action of the vector field
$X$ if
\[
Xp=Xq=0.
\]
These are precisely the string equations  appearing in the
non--scaling limit analysis performed in \cite{bx} for the Hermitian
one--matrix model with arbitrary potential, being the specific heat the first
non--trivial conserved density $2h_2=-pq$ of the AKNS hierarchy and
$p=\exp(s)$ and $q=-u\exp(-s)$ with $u=R$ and $S=\partial_1s$.  The
corresponding topological field theory is very close to the Witten's ${\Bbb
C}P^1$--sigma model coupled with topological gravity, see \cite{bx}.

Let us introduce the  complexified version of the continuous one--dimensional
Heisenberg spin chain or simply
the Heisenberg ferromagnet equation. The r\^ole of the spin field is played by
a vector field $S$ depending on $t_1,t_2$ with $S(t_1,t_2)\in{\frak sl}(2,\Bbb
C)$ such that $-\text{det}S=1$.
The Heisenberg ferromagnet equation is
\[
4\partial_2S=[S,\partial_1^2S].
\]
As is well known, see \cite{ft} and references therein,
this equation is equivalent to the  AKNS $t_2$--flow. We can write
\[
S=\text{Ad}a H
\]
where $\{E,H,F\}$ is the standard Weyl basis for ${\frak sl}(2,\Bbb C)$.
Then, the solutions of the equation
\begin{equation}
\partial_1 a\cdot a^{-1}+\text{Ad}a(pE+qF)=0\label{transS}
\end{equation}
provides solutions to the AKNS $t_2$--flow.

 The Heisenberg ferromagnet hierarchy, constructed similarly
to the AKNS hierarchy, for the spin field $S$  is the following set of
compatible equations
\[
\partial_n S=[S,S_{n+1}]
\]
with the recurrence relations
\[
\partial_1S_n=[S,S_{n+1}].
\]
For example $S_0=0, S_1=S, S_2=[S,\partial_1S]/4$. The $t_0$ flow is trivial
$\partial_0S=0$.

 If $Q_n=p_nE+h_nH+q_nF$ then
$a$ satisfies the evolution equations
\begin{equation}
\partial_na\cdot a^{-1}+\text{Ad}aQ_n=0,\label{evoa}
\end{equation}
notice that $\text{Ad}aQ_n=S_{n+1}$.

Observe that the Heisenberg ferromagnet hierarchy is invariant under the
adjoint action of $SL(2,\Bbb C)$. Given a spin field $S$ then
$\text{Ad}a_0 S$, with $a_0\in SL(2, \Bbb C)$, is also a solution.
Thus, we have an action of $PSL(2,{\Bbb C})=SL(2,{\Bbb
C})/\{\text{id},-\text{id}\}$ in the space of solutions.
In fact, when one considers ${\Bbb C}^3\cong{\frak sl}(2,\Bbb C)$ and the
Cartan--Killing bilinear form
  $(X,Y)=1/2\text{Tr}XY$,
the action above can be understood as an $SO(3,{\Bbb C})\cong
PSL(2,\Bbb C)$ action. This is the complex version of the
well known isotropy property of the $SO_3$--Heisenberg ferromagnet
($PSU_2\cong SO_3$). The local symmetries of the Heisenberg ferromagnet
are the translations and the scaling transformation.

A solution $S$
to this hierarchy gives through (\ref{transS}) a solution
$(p,q)$ to the AKNS hierarchy.
We introduce the vector field
\[
Y:=\boldsymbol{\varsigma}+\widetilde{\boldsymbol{\vartheta}},
\]
where
\[
\widetilde{\boldsymbol{\vartheta}}:=
\sum_{n\geq 0}\theta_n\partial_{n+1}.
\]

{}From Eqs.(\ref{akns},\ref{evoa}) and
\[
\partial_nS=[\partial_na\cdot a^{-1},S]
\]
one deduces
\begin{lem}
 The following equation holds
\[
YS= {\rm Ad}a(Xp\,E+Xq\,F),
\]
being $S,p,q$ connected by (\ref{transS}).
\end{lem}
from where it follows
\begin{pro}
The solution $(p,q)$ of the AKNS hierarchy satisfies the string equations
\[
Xp=Xq=0
\]
if and only if the corresponding spin field $S$ satisfies
\[
YS=0.
\]
\end{pro}

Now, we consider the scaling self--similar condition in the AKNS hierarchy and
its representation in terms of the Heisenberg ferromagnet hierarchy.
{}From (\ref{transS}) one easily gets the relation
\[
(\varsigma_{\sigma}^*a)^{-1}\cdot(\partial_1\varsigma_{\sigma}^*a)+e^{\sigma}(\varsigma_{\sigma}^*p\,E+\varsigma_{\sigma}^*q\,F)=0.
\]
 So that the solutions $(p,q)$ of the AKNS hierarchy are
scaling self--similar if and only if
\[
\varsigma_{\sigma}^*a=\pm\exp(\sigma A)\cdot a,
\]
with $A\in {\frak sl}(2,\Bbb C)$, or iff
\[
\varsigma_{\sigma}^*S=\text{Ad}\exp(\sigma A)\,S.
\]
\begin{pro}
The solution $(p,q)$ of the AKNS hierarchy is   self--similar under scaling
transformations if and only if the corresponding solution to the Heisenberg
ferromagnet hierarchy satisfies the string equation
 \[
(\boldsymbol{\varsigma}-{\rm ad A})S=0,
\]
for some $A\in{\frak sl}(2,\Bbb C)$.
\end{pro}

\end{document}